\documentclass[prb,preprint,superscriptaddress,showpacs]{revtex4}
\usepackage{graphicx}
\begin{document}
\bibliographystyle{apsrev}

\title{
Isospin phases of vertically coupled double quantum rings
under the influence of perpendicular magnetic fields}

\author{M. Royo}
\affiliation{Departament de Qu\'imica F\'isica i Anal\'itica, Universitat Jaume I,
Box 224, E-12080 Castell\'o, Spain}

\author{F. Malet}
\email{fmalet@ecm.ub.es}
\affiliation{Departament ECM and IN$^2$UB, Facultat de F\'{\i}sica,
Universitat de Barcelona. Diagonal 647, 08028 Barcelona, Spain}

\author{M. Barranco}
\affiliation{Departament ECM and IN$^2$UB, Facultat de F\'{\i}sica,
Universitat de Barcelona. Diagonal 647, 08028 Barcelona, Spain}

\author{M. Pi}
\affiliation{Departament ECM and IN$^2$UB, Facultat de F\'{\i}sica,
Universitat de Barcelona. Diagonal 647, 08028 Barcelona, Spain}

\author{J. Planelles}
\email{planelle@exp.uji.es}
\affiliation{Departament de Qu\'imica F\'isica i Anal\'itica, Universitat Jaume I,
Box 224, E-12080 Castell\'o, Spain}

\date{\today}

\begin{abstract}
Vertically coupled double quantum rings submitted to a perpendicular
magnetic field $B$ are addressed within the local spin-density functional
theory. 
We describe the structure of quantum ring molecules containing
up to 40 electrons considering different inter-ring distances
and intensities of the applied magnetic field.
When the rings are quantum mechanically strongly coupled, only 
bonding states are occupied and the addition spectrum of the
artificial molecules resembles that of
a single quantum ring, with some small differences appearing as 
an effect of the magnetic field. Despite the latter has the tendency to
flatten the spectra, in the strong coupling limit
some clear peaks are still found even when $B\neq 0$ that can be
interpretated from the single-particle energy levels analogously 
as at zero applied field, namely in terms of
closed-shell and Hund's-rule configurations.
Increasing the inter-ring distance, the occupation of the first 
antibonding orbitals washes out such structures and the addition
spectra become flatter and irregular. In the weak
coupling regime, numerous isospin oscillations are found as a
function of $B$.
\end{abstract}

%

\maketitle

\section{Introduction}

Systems made of correlated electrons confined in
semiconductor nanoscopic dot and ring structures, so-called quantum dots (QDs)
and rings (QRs) respectively, have been the subject of intense theoretical and
experimental research, see e.g. Refs. \onlinecite{Jac98,Lip08} and references therein.
From the latter point of view, for quantum dots it has been proved \cite{Tar96} the
possibility to tune over a wide range the number of electrons contained in the system,
as well as to control both the size and the shape of the dots by means of external gate
voltages, goal that has not been achieved yet for ring geometries
due to the higher complexity of their fabrication process, 
\cite{Gar97,Lor00,Fuh01} which involves several experimental techniques
such as atomic force microscopy,\cite{Hel99} strain-induced self-organization \cite{Gar97}
or droplet molecular beam epitaxy. \cite{Gon05}

The interest of QRs arises from their peculiar behavior in the presence of a perpendicularly
applied magnetic field ($B$), which is very distinct from that observed in QDs and shows up as
an oscillatory behavior of their energy levels as a function of $B$.
This property, together with the fact that in narrow enough QRs the electrons experiment a
nearly one-dimensional Coulomb repulsion, leads to the integer and fractional
Aharonov-Bohm effects, usually associated with the appearance of the so-called persistent currents
in the ring. \cite{Kle07} These quantum-interference phenomena have been experimentally reported
\cite{Ihn05} and have motivated a series of theoretical works whose number is steadily
increasing,
see e.g., Refs. \onlinecite{Cha94,Kri94,Emp99,Lin01,Aic06,Liu08} and references
therein. 

One of the most appealing possibilities offered by electron systems confined in semiconductor
heterostructures is their ability to form coupled entities, usually referred to as ``artificial
molecules'', in which the role of the constituent ``atoms'' is played by single quantum dots or
rings and that have analogies with natural molecules such as the hybridization of the
electronic states forming molecular-like orbitals. In addition, these artificially coupled
systems present important advantages such as a tunable ``interatomic'' coupling by means, e.g.,
of the modification of the relative position/size of the constituents.
This fact has, besides its intrinsic interest, potential relevance to quantum information
processing schemes since basic quantum gate operations require controllable coupling between
qubits. In this sense, artificial molecules based on two coupled QDs, called quantum dot
molecules (QDMs) have been proposed
as scalable implementations for quantum computation purposes and have received great attention
from the scientific community in the last years --see e.g. Refs. \onlinecite{Bli96,Sch97,Ron01,
Hol02,Ota05,Par00,Pal95,Anc03,Aus04,Bel06} and references therein.

Also, molecular-beam epitaxy techniques have recently allowed the synthesis of
{\it quantum ring molecules} (QRMs) in the form of concentric double QRs \cite{Man05,Kur05} and 
vertically stacked layers of self-assembled QRs, \cite{Sua04,Gra05} the optical and structural 
properties of the latters having also been characterized by photoluminescence spectroscopy and 
by atomic force microscopy, respectively.
This has sparked theoretical studies on the structure and optical response of both
vertically and concentrically coupled QRs of different complexity and scope, revealing properties
different from those of their dot counterparts due to the non-simply connected ring topology.
For instance, studies on the single-electron spectrum of vertical QRMs \cite{Ahn00,Li04} have shown 
that the electronic structure of these systems is more sensitive to the inter-ring distance than that 
of coupled QDs. As a consequence, in ring molecules quantum tunneling effects are enhanced since less 
tunneling energy is required to enter the molecular-type phase. Also, the consideration
of ``heteronuclear'' artificial molecules constituted by slightly different QRs offers the interesting
possibility to control the effective coupling of direct-indirect excitons \cite{Dia07}
by means of the application of a magnetic field and taking advantage of the fact that charge
tunneling between states with distinct angular momentum is strongly suppressed by orbital selection
rules. To this end, some authors have considered the case of QRMs made of strictly one-dimensional,
zero-thickness QRs and have used diagonalization techniques to address the few-electron
problem. \cite{Ahn00,Sza07,Dia07,Sza08}
The simultaneous effect of both electric and magnetic fields applied to a single-electron QRM has
also been studied \cite{Pia07} --see also Ref. \onlinecite{Ahn00}-- and the optical response of 
QRMs where the thickness of the constituent QRs is taken into
account has been obtained.\cite{Cli05} In addition, the spatial correlation between electron pairs
in vertically stacked QRs only electrostatically coupled has been shown to undergo oscillations 
as a function of the magnetic flux,
with strongly correlated situations between ground states with odd angular momentum 
turning out to occur even at large inter-ring distances.
\cite{Sza07} More recently, the structure of a QRM made of two vertically stacked quantum
rings has been addressed at zero magnetic field for a few tens of electrons within 
the local spin-density functional theory (LSDFT) neglecting\cite{Cas06} and incorporating\cite{Mal06}
the vertical thickness of the constituent QRs.

In this work we address the ground state (gs) of two thick, vertically coupled 
identical quantum rings forming ``homonuclear'' QRMs populated with up to 40 electrons and
pierced by a perpendicularly applied magnetic field.
We extend in this way our previous study,\cite{Mal06} addressing the appearance and physical
interplay between the spin and isospin\cite{Pal95} degrees of freedom as a function of
the variation of both the intensity of the magnetic field and the inter-ring separation.
Modelling systems charged with such large number of electrons requires the employment of
methodologies that minimize the computational cost.
Here we have made use of the LSDFT,\cite{Emp99,Aic06} whose accuracy for the
considered values of the magnetic
field has been assessed\cite{Anc03} by comparing the obtained results for a single QD
with those given by the current-spin-density functional theory (CSDFT),\cite
{Fer94} which in principle is better-suited for high magnetic fields, and also with
exact results for artificial molecules.\cite{Ron99}

This paper is organized as follows.
In Sec. II we briefly introduce the LSDFT and the model used to represent the vertical QRMs.
In Sec. III we discuss the obtained results for some selected configurations,
and a summary is given in Sec. IV.

\section{Density functional calculation for many-electron vertical
quantum ring homonuclear molecules}

The axial symmetry of the system allows one to work in cylindrical
coordinates. The confining potential $V_{cf}(r,z)$ has been taken
parabolic in the $xy$-plane with a repulsive core around the origin, plus
a symmetric double quantum well in the $z$-direction, each one with width 
$w$, depth $V_0$, and separated by a distance $d$. 
To improve on the convergence of the calculations, the double-well
profile has been slightly rounded off, as illustrated
in Fig. 2 of Ref. \onlinecite{Anc03}.
The potential thus reads $V_{cf}(r,z)=V_r(r)+V_z(z)$, where
\begin{eqnarray}
V_r(r) &=&
V_0 \,\Theta(R_0-r) +
\frac{1}{2}\, m \,\omega_0^2\, (r - R_0)^2 \, \Theta (r-R_0) 
\nonumber
\\
V_z(z)&=&V_0\left\{
\begin{array}{ll}
\frac{1}{1+e^{(z+d/2+w)/\sigma}}-
\frac{1}{1+e^{(z+d/2)/\sigma}}
& \; {\rm if} \; z \le 0\\
\frac{1}{1+e^{(z-d/2)/\sigma}}-
\frac{1}{1+e^{(z-d/2-w)/\sigma}}
& \; {\rm if} \; z > 0 \; , \\
\end{array}\right. 
\label{eq2}
\end{eqnarray}
with $\sigma=2\times 10^{-3}$ nm, and $\Theta(x)=1$ if $x>0$ and zero
otherwise.
The convenience of using a hard-wall confining potential to describe the 
effect of the inner core in QRs is endorsed by several works in the 
literature.\cite{Li01} We have taken $R_0=10$ nm, $V_0=$ 350 meV, 
$\hbar \omega_0=6$ meV and $w=5$ nm.
These parameters determine the confinement for the electrons together with
the distance between the constituent quantum wells that is varied to study
QRMs in different inter-ring coupling regimes.

For small electron numbers ($N$), it is justified to take $\omega_0$
to be $N$-independent. However, in a more realistic scheme its value
should be tuned according to the number of electrons contained in the
system, relaxing the confinement as the latter is increased.
In the case of quantum dots it has oftenly been used a $N^{-1/4}$-dependence
that arises from the $r$-expansion near the origin of the Coulomb potential
created by a two-dimensional uniform positive charge distribution --jellium
model-- and that it is generalized to the case of quantum dot molecules as
$\omega_0 = \kappa N^{-1/4}_{B}$, $N_B$ being the
number of electrons filling bonding orbitals --see below. The rationale
for this generalization is given in Ref. \onlinecite{Aus04}.
It is clear that the mentioned $N$-dependence would be harder to justify
for QRs, and in fact no alternative law is known for a single QR that could
be generalized to the case of QRMs. For this reason, in this work we have taken
$\omega_0$ to be $N$-independent, which is to some extend less
realistic for the largest values of $N$ we have considered.

Considering the $N$-electron system placed in a magnetic field parallel
to the $z$-axis, within the LSDFT in the effective mass, dielectric constant
approximation, the Kohn-Sham equations \cite{Pi01,Anc03} in cylindrical coordinates
read
\begin{eqnarray}
& & \left[-\frac{1}{2} \left( \frac{\partial^2}{\partial r^2}
+ \frac{1}{r} \frac{\partial}{\partial r} - \frac{l^2}{r^2}
+ \frac{\partial^2}{\partial z^2} \right) - \frac{\omega_c}{2}\,l
+ \frac{1}{8} \omega_c^2 r^2 + V_{cf}(r,z) \right.
\nonumber
\\
& &
\label{eq1}
\\
&+& \left. V_H + V_{xc} + \left( W_{xc}
+\frac{1}{2} g^* \mu_B B\right) \eta_{\sigma} \right]
u_{n l \sigma}(r,z) =
\varepsilon_{n l \sigma} u_{n l \sigma}(r,z) \,\, ,
\nonumber
\end{eqnarray}
where the single-particle (sp) wave functions have been taken to be
of the form $\phi_{n l\sigma}(r,z,\theta,\sigma)=
u_{n l\sigma}(r,z) e^{-\imath l \theta} \chi_{\sigma}$ with
$n =0, 1, 2, \ldots$, $l =0, \pm 1, \pm 2, \ldots$, $-l$ being
the projection of the single-particle orbital angular momentum
on the symmetry axis, and $\sigma$=$\uparrow$$(\downarrow)$
representing spin-up(down) states.
The vector potential has been chosen in the symmetric gauge,
namely ${\bf A}= B (-y,x,0)/2$; $\mu_B = \hbar e/(2 m_e c)$ and
$\omega_c =e B/c$ are, respectively, the Bohr magneton and the
cyclotron frequency, and $\eta_{\sigma}$=$+1(-1)$ for
$\sigma$=$\uparrow$$(\downarrow)$;
$V_H(r,z)$ is the direct Coulomb potential, and $V_{xc}={\partial
{\cal E}_{xc}(n,m)/\partial n}\vert_{gs}$ and
$W_{xc}={\partial {\cal E}_{xc}(n,m)/\partial m}\vert_{gs}$
are the  variations of the exchange-correlation
energy density ${\cal E}_{xc}(n,m)$ in terms of the electron
density $n(r,z)$ and of the local spin-magnetization
$m(r,z)\equiv n^{\uparrow}(r,z)-n^{\downarrow}(r,z)$ taken at the gs.
${\cal E}_{xc}(n,m) \equiv {\cal E}_{x}(n,m) + {\cal E}_{c}(n,m)$
has been built from three-dimensional homogeneous electron gas
calculations; this yields a well-known,\cite{Lun83} simple analytical
expression for the exchange contribution ${\cal E}_{x}(n,m)$.
For the correlation term ${\cal E}_{c}(n,m)$ we have used
the parametrization proposed by Perdew and Zunger.\cite{Per81}
Details about how the Kohn-Sham and the Poisson equations
have been solved can be found in Ref. \onlinecite{Pi01}.
Notice the use in Eq. (\ref{eq1}) of effective atomic units
$\hbar=e^2/\epsilon=m=$1, where $\epsilon$ is the dielectric constant
and $m$ the electron effective mass. In units
of the bare electron  mass $m_e$ one has $ m = m^* m_e$,
the length unit being the effective
Bohr radius $a_0^* = a_0\epsilon/m^*$
and the energy unit the effective Hartree $H^* = H  m^*/\epsilon^2$.
In the numerical applications we have considered
GaAs quantum rings, for which  we have taken $\epsilon$ = 12.4, and
$m^*$ = 0.067; this yields $a^*_0 \sim$ 97.9 ${\rm \AA}$ and
$H^*\sim$ 11.9 meV, the effective gyromagnetic constant being $g^*=-0.44$.

To label the gs configurations (``phases'') we use an adapted version of the
ordinary spectroscopy notation,\cite{Ron99} namely $^{2S+1}L^{\pm}_{g,u}$, 
where $S$ and $L$ are the total $|S_z|$ and $|L_z|$, 
respectively. The superscript $+(-)$ corresponds to symmetric (antisymmetric)
states under reflection with respect to the $z=0$ plane bisecting the QRMs,
and the subscript $g(u)$ refers to positive(negative) parity states. 
All these are good quantum numbers even in the presence of an axial magnetic field.
By analogy with natural molecules, symmetric and antisymmetric states
are referred to as bonding (B) and antibonding (AB) orbitals, respectively.
We have defined the ``isospin'' quantum number $I_z$ --bond order in Molecular
Physics-- as\cite{Par00,Anc03,Ron99} $I_z = (N_B -N_{AB})/2$, $N_{B(AB)}$ being
the number of occupied bonding(antibonding) sp states.

\section{Results}

Due to the large number of variables needed to characterize a given
QRM configuration (electron number, magnetic
field and inter-ring distance), we limit ourselves to present
results in a limited range of values for such variables, aiming at
discussing calculations that might illustrate the appearance of
some properties of the systems under study.
For the sake of comparison, we have also addressed one single QR 
symmetrically located with respect to the $z=0$ plane with the
same thickness (5 nm) and radial confinement as the coupled rings.

\begin{figure}[f]
\centerline{\includegraphics[width=15cm,clip]{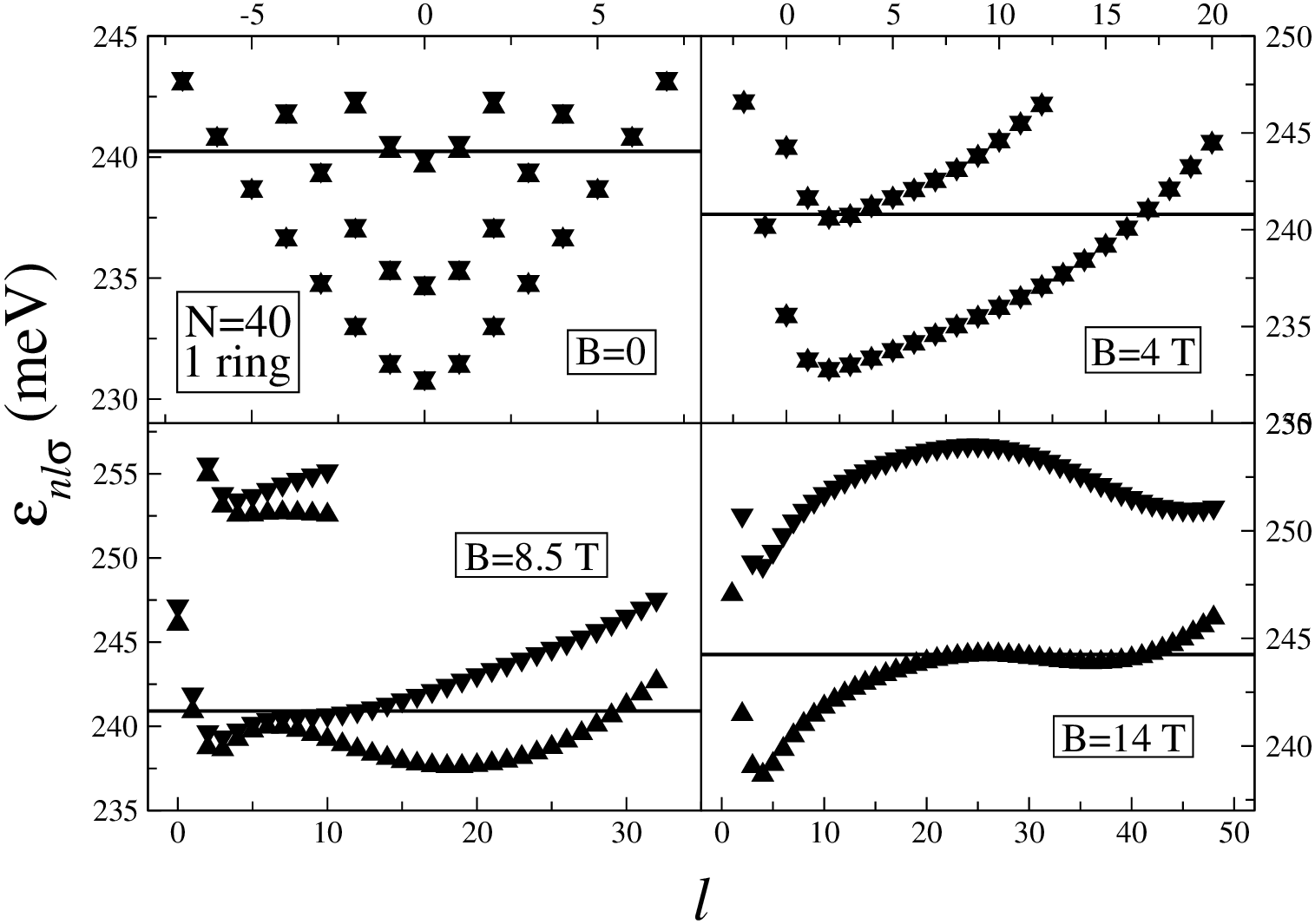} }
\caption{Single-electron energies (meV) as a function of $l$
for a $N=40$ single ring.
Upward(downward) triangles denote $\uparrow$$(\downarrow)$ spin states.
The horizontal lines represent the Fermi levels. The value of $B$ (T)
is indicated in each panel.
}
\label{fig1}
\end{figure}

Fig. \ref{fig1} shows the Kohn-Sham sp levels for one single ring hosting $N$=40
electrons as a function of $l$ for different values of the applied magnetic field.
As it is well known, these levels are $\pm l$-degenerate at $B=0$. In this particular 
case, the gs has $S_z=1$, and it is made up of symmetric (with respect to $z=0$) sp 
states with up to $n=3$. In the non-interacting single-electron model, in which
the Coulomb energy is not considered and consequently the sp wave functions
factorize into a $r$-dependent and a $z$-dependent part with associated
quantum numbers $n_r$ and $n_z$, i.e.,
$u_{nl}(r,z) \rightarrow {\cal U}_{n_r}(r) {\cal Z}_{n_z}(z)$, one would say that
the gs is made up of sp states with $n_z=0$ and radial quantum numbers up to $n_r=3$.

When $B\neq$ 0, the $\pm l$-degeneracy is lifted and, on the other hand,
the $l<0$ sp levels become progressively depopulated
in favor of those with $l>0$ as the magnetic field increases until eventually
--at about $\sim$ 4 T-- only $l>0$ orbitals are filled. At this point,
only a few states with $n=2$ are occupied, and the ring has $S_z=0$. From this
value of $B$ on, the simultaneous filling of increasingly higher-$l$ states and
those close to $l=0$ gives rise to configurations containing only states with $n=1$
and with large values of the total spin (e.g., $S_z=9$ for $B=8$ T). Eventually, the system
becomes fully spin-polarized at $B \sim$ 13.5 T. It is worth noticing the conspicuous
bending of the ``Landau bands'' (sets of bonding or antibonding states characterized
by the same $n$ and spin, and different value of $l$),
instead of displaying a fairly flat region, as it happens
when the in-plane confinement is produced by a jellium-like potential,\cite{Emp99}
but not with our present choice of a $N$-independent parabola.
It is also worth to stress that, due to the much stronger confinement in the vertical 
direction as compared to that in the radial one, only symmetric states
are occupied.

\begin{figure}[f]
\centerline{\includegraphics[width=15cm,clip]{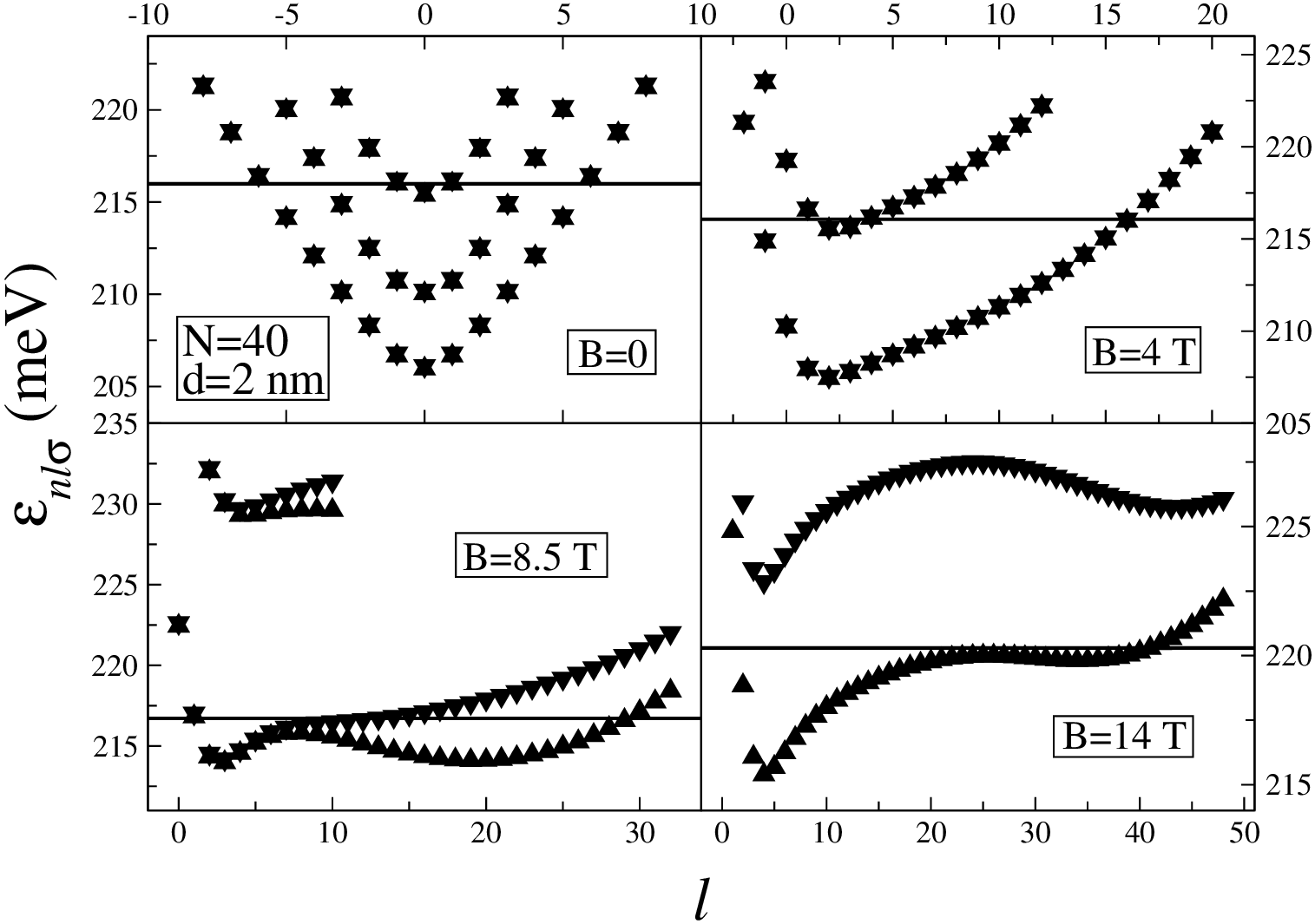} }
\caption{Same as Fig. \ref{fig1} for a QRM with $N=40$ electrons and $d=2$ nm.
Notice that due to the small separation between the rings only bonding
states are occupied, the antbonding ones lying at much higher energies.
}
\label{fig2}
\end{figure}

\begin{figure}[f]
\centerline{\includegraphics[width=15cm,clip]{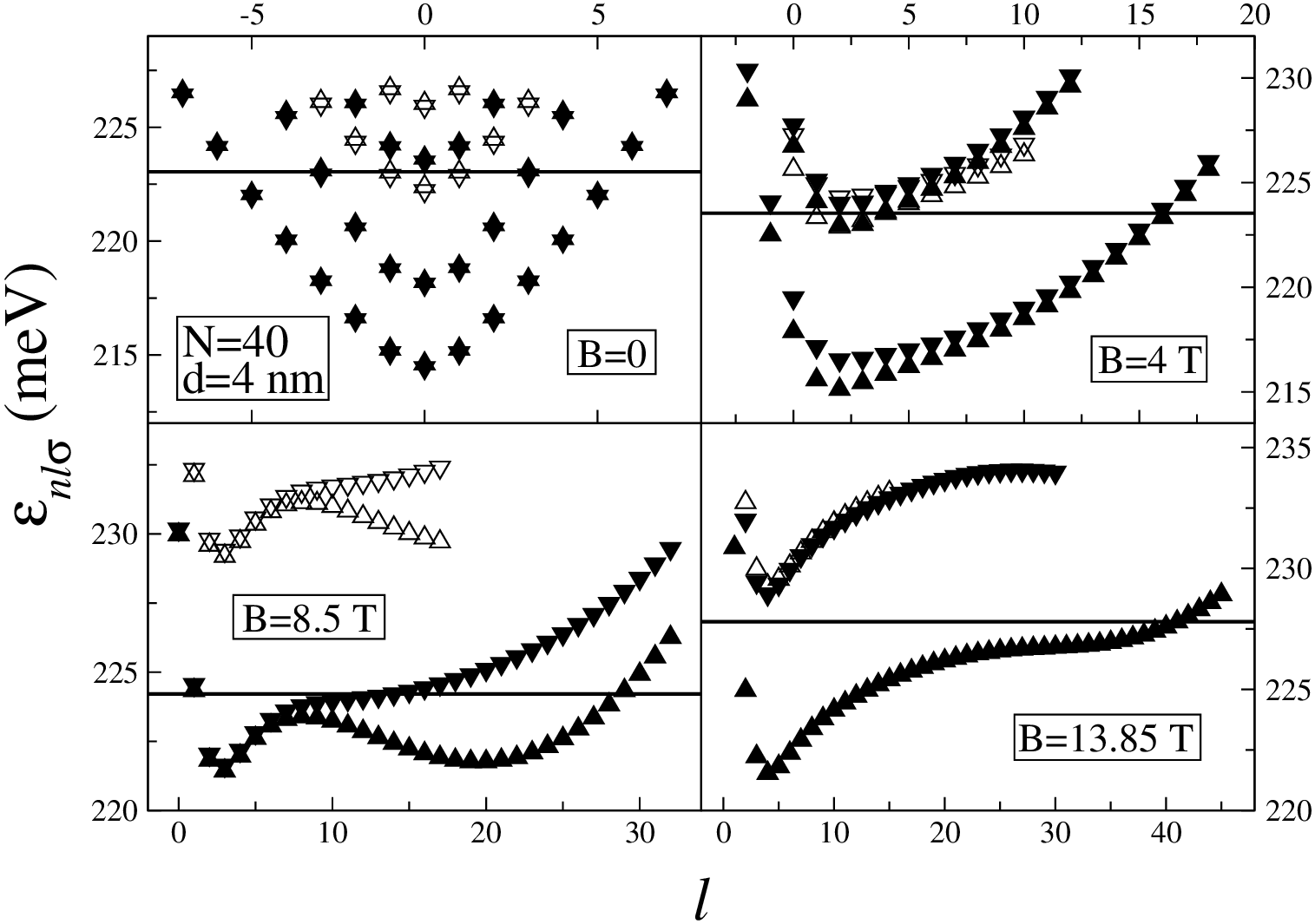} }
\caption{Same as Fig. \ref{fig2} for $d=4$ nm.
The increase of the inter-ring separation allows for
the occupation of both bonding and antibonding states, 
indicated by solid and open triangles respectively.
}
\label{fig3}
\end{figure}

\begin{figure}[f]
\centerline{\includegraphics[width=15cm,clip]{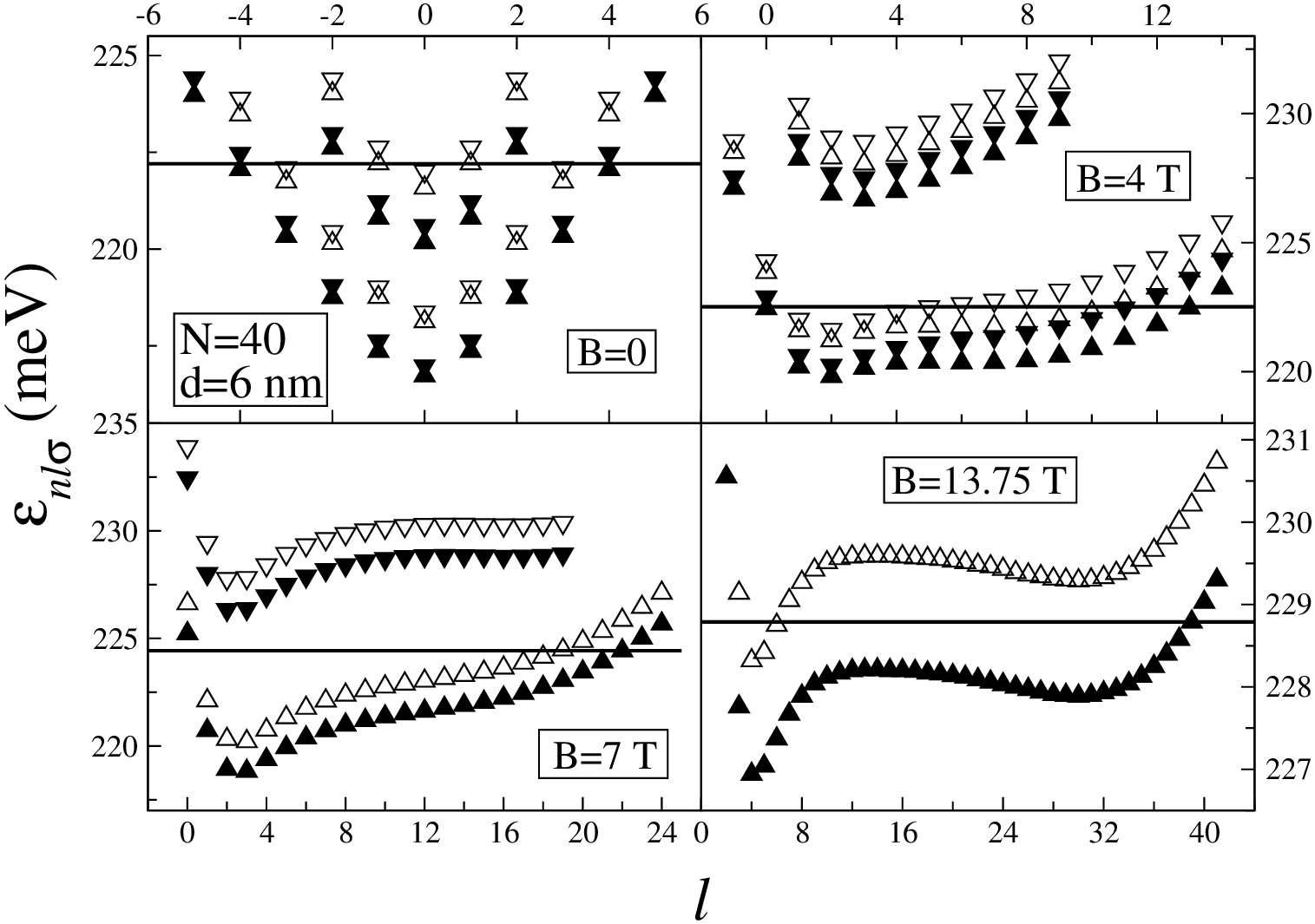} }
\caption{Same as Fig. \ref{fig3} for $d=6$ nm.
}
\label{fig4}
\end{figure}

Analogously, the energy levels corresponding to QRMs with $N=40$ and inter-ring
distances $d=2$, 4 and 6 nm are shown in Figs. \ref{fig2}-\ref{fig4}.
One can see the gradual evolution of the system 
as $d$ increases; indeed, at $d=2$ nm the spectrum is very similar to that 
of the single ring, with only bonding sp states being occupied.
As $d$ increases, a few antibonding orbitals become populated at small $B$'s, as
one can see from the top panels of Fig. \ref{fig3}, corresponding to $d=4$ nm,
but eventually for increasing values of $B$ the QRMs have again ground states where
only bonding states are populated, as can be seen from the bottom panels of the
same figure. For this inter-ring distance, the fully spin-polarized state is reached
at $B \sim$ 13.75 T. Finally,
for the largest ring separation considered, namely $d=6$ nm, a large amount
of antibonding orbitals become occupied giving rise to small $I_z$'s instead
of the fairly large isospin values found for similar configurations at smaller
distances (compare the bottom panels in Fig. \ref{fig4} with those in
Figs. \ref{fig2} and \ref{fig3}).
In particular, the fully spin-polarized gs is found at about $B\sim$ 7 T with 
$I_z=2$, whereas for $d=2$ and 4 nm it appears near $B=14$ T
and has the maximum possible isospin value, namely $I_z=20$.
At $d=6$ nm, the maximum-spin state naturally consists of two distinct bands, one
made up of bonding and another of antibonding states. These configurations are the
QRM-analogues of the maximum density droplet (MDD) configurations found for QDMs
at similar inter-dot distances, called respectively MDD$_B$ and MDD$_{AB}$ in 
Ref. \onlinecite{Aus04}.
Increasing further the magnetic field causes the progressive occupation of
higher-$l$ orbitals, which provokes the depopulation of the
antibonding band and the consequent increase of $I_z$. For the highest 
considered magnetic field ($B\sim$ 14 T), some antibonding orbitals are 
still occupied, yielding $I_z=17$.

These results are a consequence of the evolution with $d$ of the energy
difference between bonding and antibonding states, $\Delta_{SAS}$,
which accurately varies as a function of the inter-ring distance
according to the law $\Delta_{SAS}= \Delta_0 e^{-d/d_0}$, already found
for QDMs.\cite{Par00}
In our case, from the difference in energy of single-electron (bonding and
antibonding) QRMs we have obtained $\Delta_0=82$ meV and $d_0=1.68$ nm,\cite{Mal06}
values which have turned out to be unaffected by the applied magnetic field.
Clearly, the value of $\hbar \omega_0$ as compared to $\Delta_{SAS}$, which allows 
to discern between the strong ($\hbar \omega_0 \lesssim \Delta_{SAS}$) and the weak
($\hbar \omega_0 \gg \Delta_{SAS}$) quantum mechanical coupling regimes,
has a crucial influence on the actual filling of bonding
and antibonding sp states at a given inter-ring distance. Indeed, increasing
$\omega_0$ while keeping constant the double well structure may favor the population
of antibonding orbitals for large enough values of $N$. \cite{Anc03} This can be
understood from the non-interacting electron model, in which the single-electron
energies are the sum of two independent terms, one arising from the $z$-localization
and characterized by the quantum number $n_z$, and another, which increases as 
$\omega_0$ does, arising from the $r$-localization and depending on $l$ and the 
radial quantum number $n_r$. If $N$ is large enough,
the QRMs can minimize its energy by populating antibonding states with
low values of $n_r$ and $l$ instead of going on populating bonding states
with higher quantum numbers. This explains why some antibonding states were
filled even for $d=2$ nm at $B=0$ and $N=40$ for the QRMs of Ref. \onlinecite{Mal06},
where $\omega_0$ was taken to be 15 meV, value almost three times larger than the
one considered in the present work. 
\cite{Anc03}

This particular structure of the bonding and antibonding bands at high magnetic fields
may have some observable effects on the far-infrared response of QRMs. Indeed, 
since the dipole operator cannot connect bonding with antibonding sp states,
for QRMs in the weak coupling limit one would expect the dipole spectrum to
display additional fragmentation in the characteristic edge modes of the ring
geometry \cite{Emp99} due to the contribution of the antibonding electron-hole 
pairs (see e.g. the bottom panels of Fig. \ref{fig4}).

\begin{figure}[f]
\centerline{\includegraphics[width=12cm,clip]{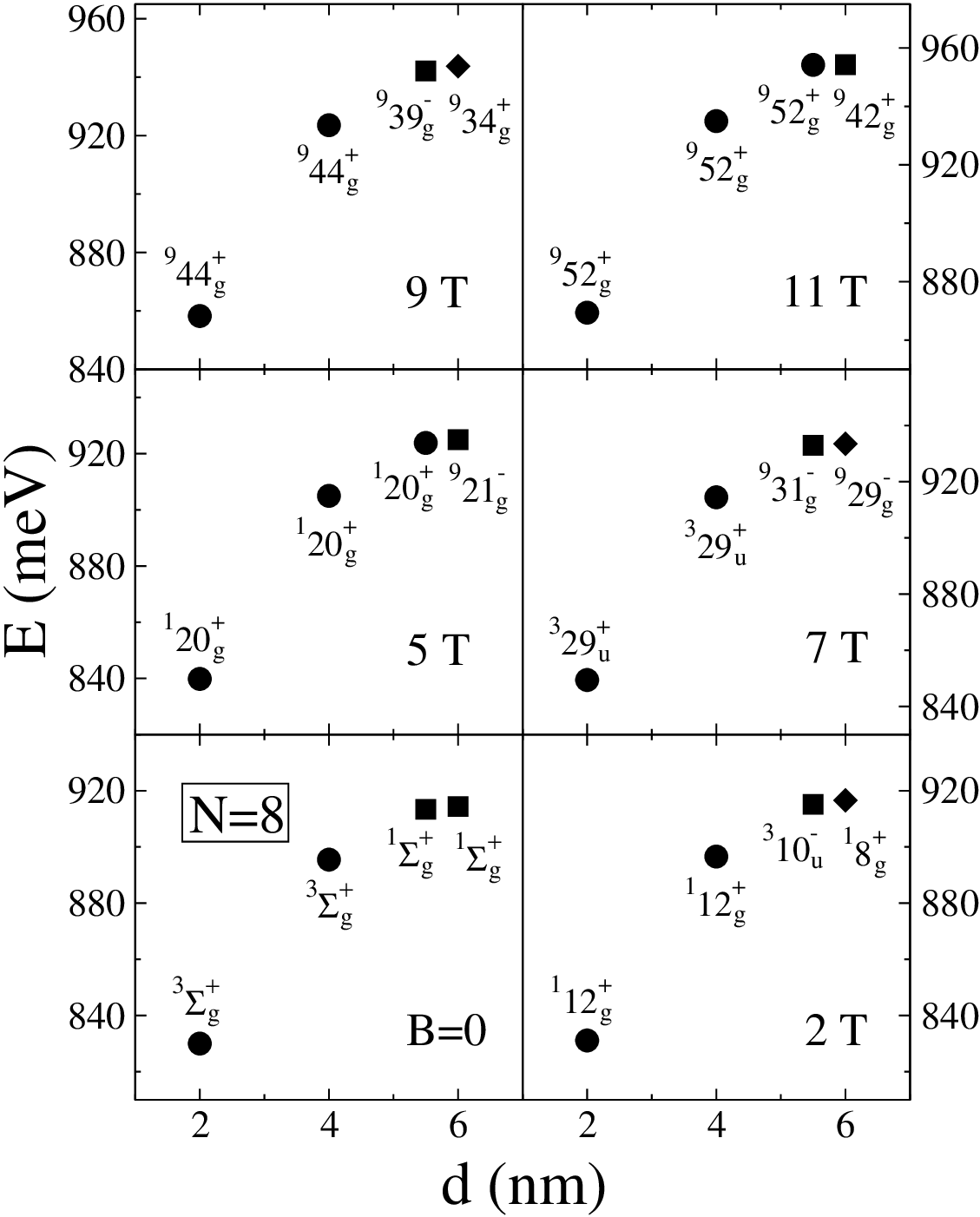} }
\caption{
Energy (meV) and gs molecular phases of the QRM with $N=8$ electrons
as a function of the inter-ring distance for different values of $B$.
For a given $B$ value, different phases are
represented by different symbols.
}
\label{fig5}
\end{figure}

Figure \ref{fig5} shows the evolution with $d$ of the gs energy and the
molecular phase of QRMs made up of $N=8$ electrons and submitted to
magnetic fields of different intensities.
Notice that even moderate values of $B$ give rise to ground states with 
large total angular momentum, which increases as the magnetic
field does. For this reason, we have denoted it by its actual value instead
of employing the usual notation with upper Greek letters, except for the 
cases with $L_z=0$.
Similar conclusions can be drawn for all the values of the magnetic field
we have considered; on the one hand, for the studied inter-ring distances, 
the energy of the molecular phases increases with $d$ due to the enhancement
of the energy of the bonding states, \cite{Pi01} which dominates over the 
decrease of the Coulomb energy --for larger distances the constituent QRs are 
so apart that eventually this decrease dominates and the tendency is reversed.
On the other hand, one can see that the first phase transitions are always 
found at the largest inter-ring distances since, as happens for QDMs in the 
few-electron limit, they are due to the replacement of an occupied bonding sp 
state by an empty antibonding one. This also explains why in most of the cases,
and especially for the highest magnetic fields, the total angular momentum of
the QRMs in the weakest coupling regime is reduced: the filled antibonding
orbitals have lower $l$'s than the replaced bonding states.

\begin{figure}[f]
\centerline{\includegraphics[width=12cm,clip]{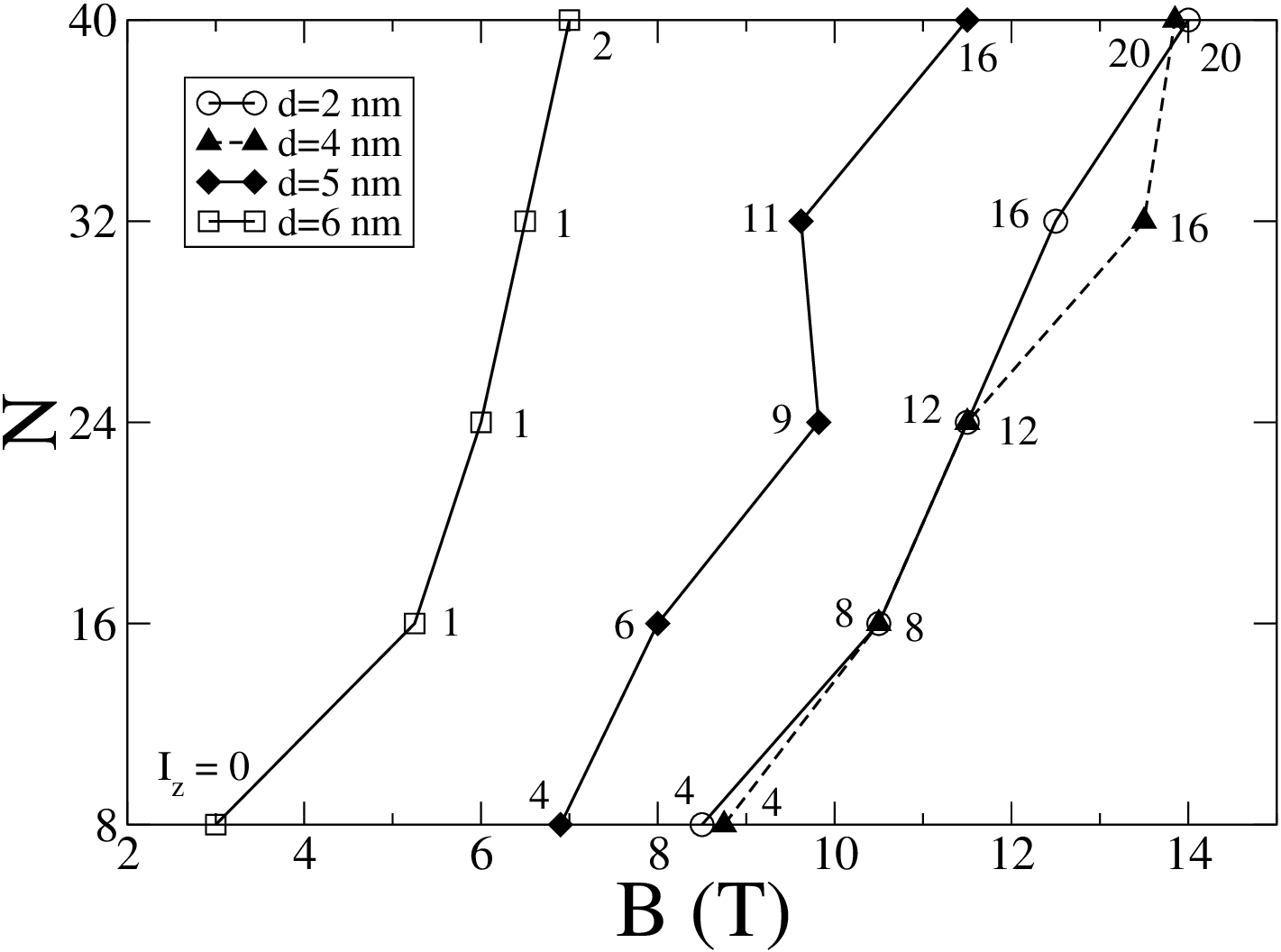} }
\caption{
$B$ values at which the selected QRMs become fully spin-polarized for
the inter-ring distances $d=2, 4$, and 6 nm. The isospin
of each configuration is also indicated. The lines have been drawn
to guide the eye.
}
\label{fig6}
\end{figure}

We have determined the magnetic field that gives rise to 
ring molecules with fully spin-polarized gs, and show it
in Fig. \ref{fig6} as a function of $N$
for different inter-ring distances going from the strong
to the weak quantum mechanical coupling regimes. 
The isospin value of each configuration is also indicated.
The number of electrons, $N= 8\times M$
with $M=1$ to 5,  was chosen with the aim of
producing closed-shell structures at $B=0$ in the weak coupling limit.
One can see that the results for $d=2$ and 4 nm are very close,
with only noticeable differences for $N=32$. This can be
understood from the bottom panels of Figs. \ref{fig2} and \ref{fig3}, which
show that for rather large magnetic fields only bonding orbitals are
occupied for both ring separations.
Contrarily, from Fig. \ref{fig4} one can see that in weaker coupling regimes
the filling of antibonding states favors the fully spin-polarization of the
QRMs at low $B$ intensities as compared to those needed when the rings are
closer to each other, which explains the differentiated results corresponding
to $d=5$ and 6 nm in Fig. \ref{fig6}. 

\begin{figure}[f]
\centerline{\includegraphics[width=10cm,clip]{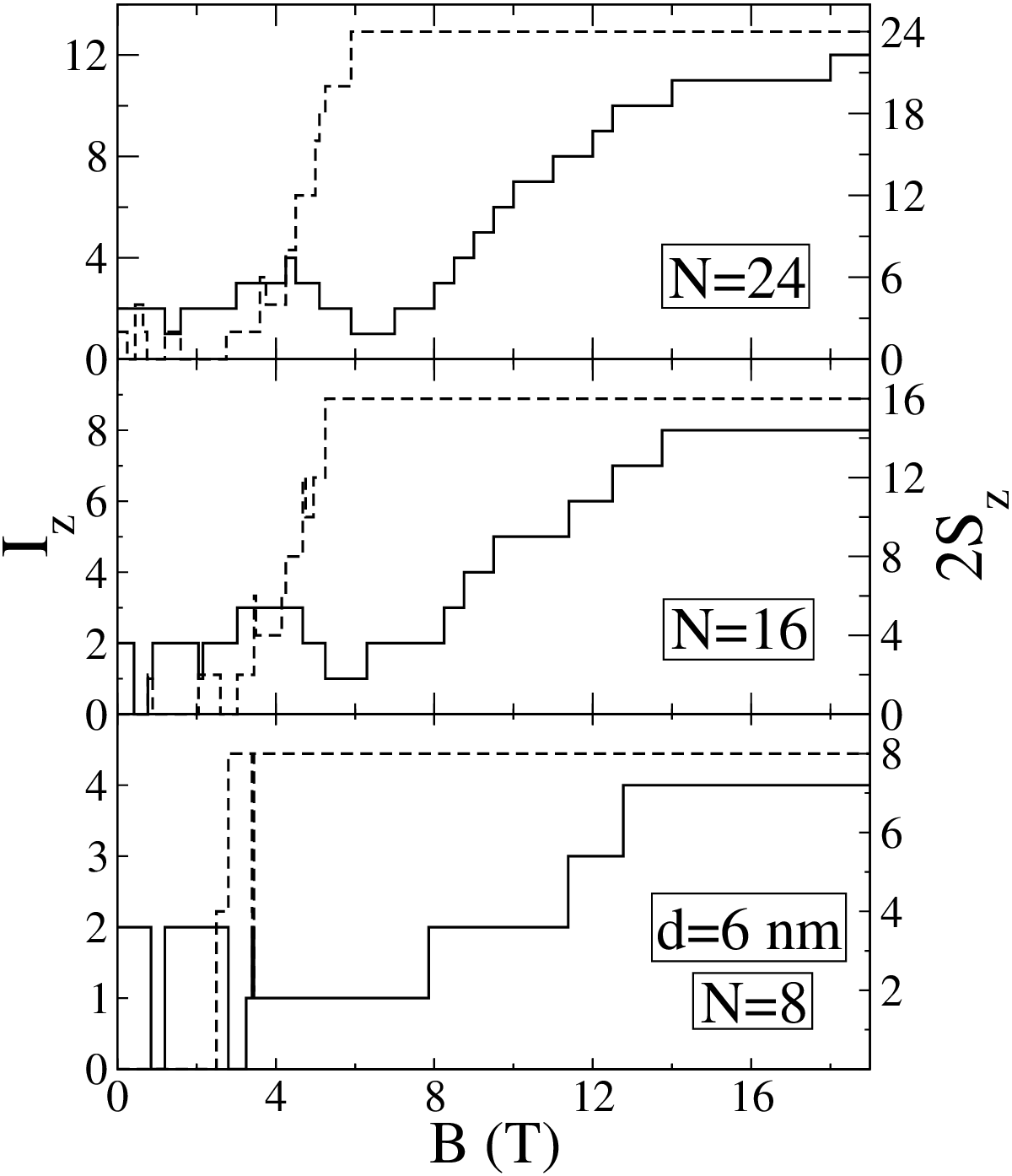} }
\caption{
Isospin (solid line, left scale) and spin (dotted line, right scale)
values as a function of $B$ for the QRMs with $d=6$ nm and $N=8,16$
and 24.
}
\label{fig7}
\end{figure}

When antibonding orbitas are populated, the variation of the magnetic field 
yields numerous transitions between different molecular phases with different 
isospin that are more complex
than these observed in vertically coupled QDs. This particular behavior
is mainly due to the periodic destabilization suffered by the lowest-$l$
occupied orbitals induced by the magnetic field, which is a direct
consequence of the Aharonov-Bohm effect and makes it rather difficult
to find a pattern among the observed evolutions for the different electronic
populations.
The spin and isospin phases as a function of the magnetic field are shown in
Fig. \ref{fig7} for $d=6$ nm corresponding to $N=8, 16,$ and 24.
It can be seen that in all cases at $B=0$ the QRMs have $I_z=2$
and $S_z=1$; when $B$ is increased, non-monotonic spin and isospin oscillations
with $\Delta I_z=\pm 1$ and $\Delta S_z=\pm 1$ and 2 appear, respectively.
Two facts, also present in QDMs, \cite{Anc03,Aus04} are worth to be stressed:
on the one hand, molecular phase changes from $-$($+$) to $+$($-$) ground states 
--recall that, as explained in Sec. II, this sign is related to the symmetry of 
the molecular configuration-- involve $\Delta I_z=+1$(-1) flips;
on the other hand, quite often the transitions 
in both magnitudes take place simultaneously, except obviously when the QRMs 
reach the full spin-polarization, point from which on the isospin increases in 
one-unit jumps until the system is made up of only bonding states.

\begin{figure}[f]
\centerline{\includegraphics[width=14cm,clip]{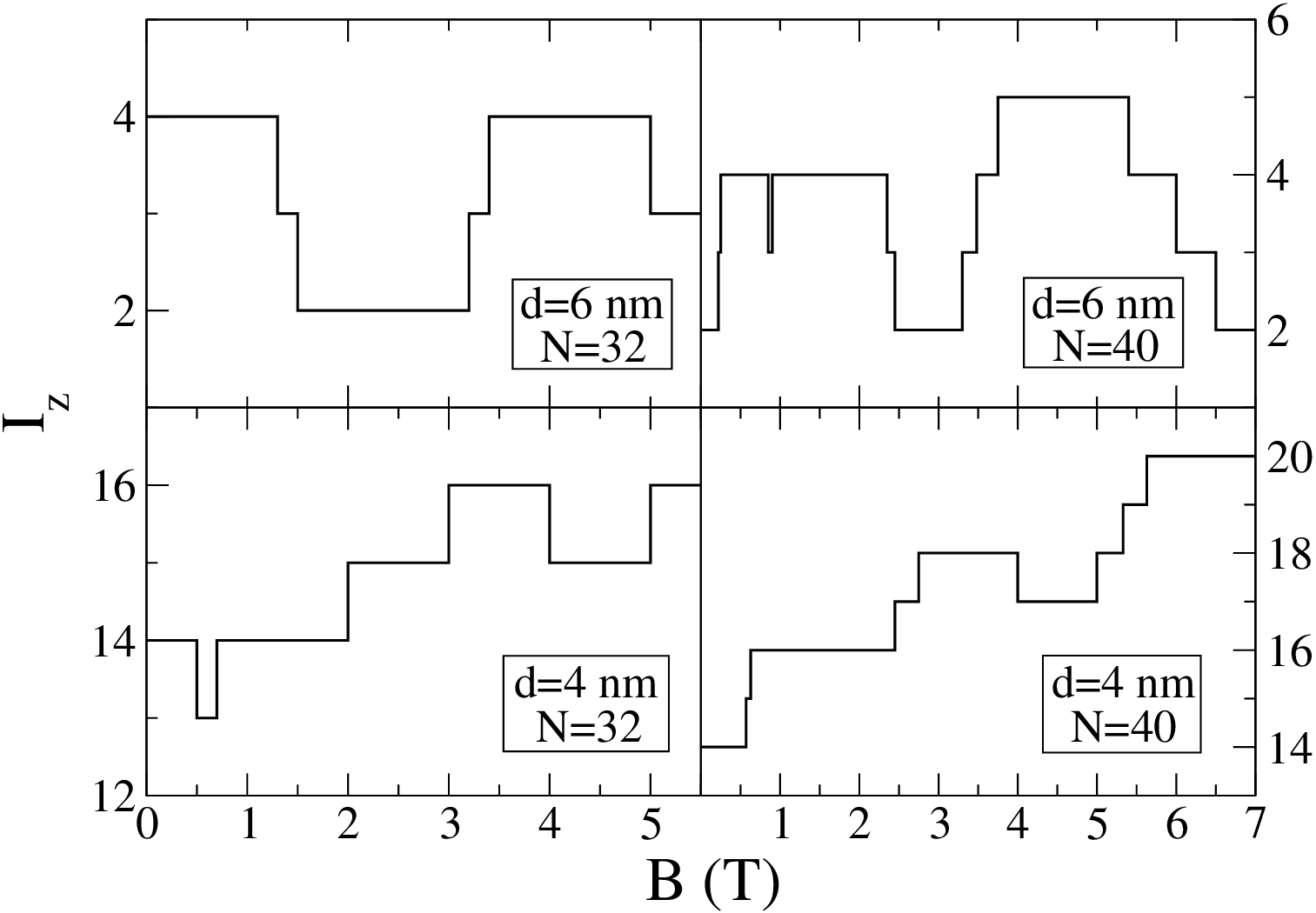} }
\caption{
Isospin values as a function of $B$
for the QRMs with $d=4$ and 6 nm, and $N=32$ and 40.
}
\label{fig8}
\end{figure}

The comparison of the isospin phases for QRMs with $d=4$ and 6 nm is 
presented in Fig. \ref{fig8} for $N=32$ and 40. Clearly, the highest
values of $I_z$ appear for the smallest inter-ring distances, as
expected from the single-particle levels shown in Fig. \ref{fig3}, 
corresponding to $d=4$ nm and $N=40$, in which only a few antibonding 
orbitals are occupied for low values of $B$. Indeed, one can see from
the bottom panels of Fig. \ref{fig8} that for this inter-ring distance
magnetic fields of about 5 T are enough to yield configurations with
the maximum isospin value $N/2$, whereas for the QRMs with $d=6$ nm
such values of $B$ still correspond to small $I_z$'s due to the large
amount of filled antibonding states.

\begin{figure}[f]
\centerline{\includegraphics[width=10cm,clip]{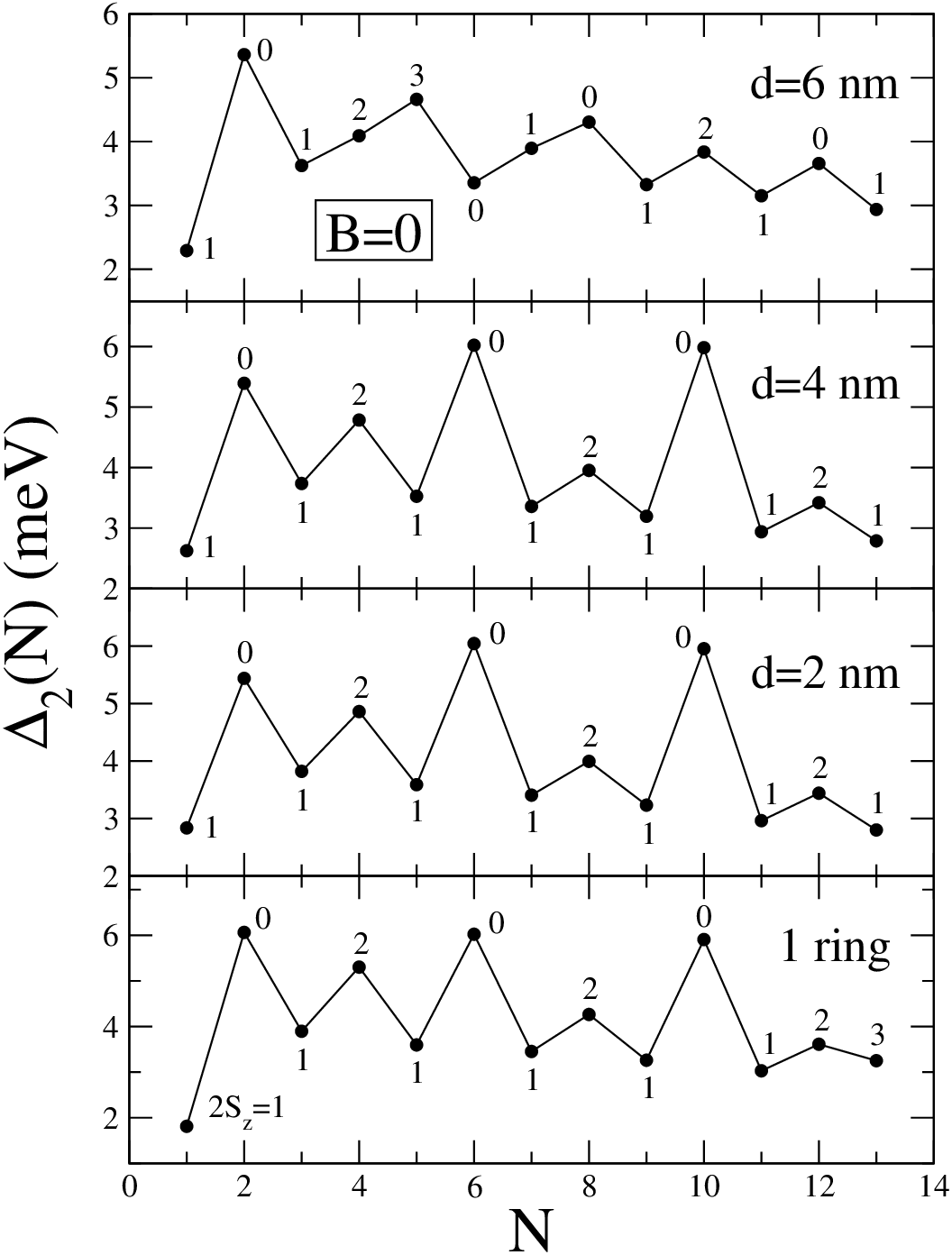} }
\caption{
Addition energies for QRMs at $B=0$ and inter-ring distances $d=2, 4$,
and 6 nm, and for a single QR (bottom panel). The value of $2S_z$
corresponding to each $N$ is also indicated.
}
\label{fig9}
\end{figure}

\begin{figure}[f]
\centerline{\includegraphics[width=10cm,clip]{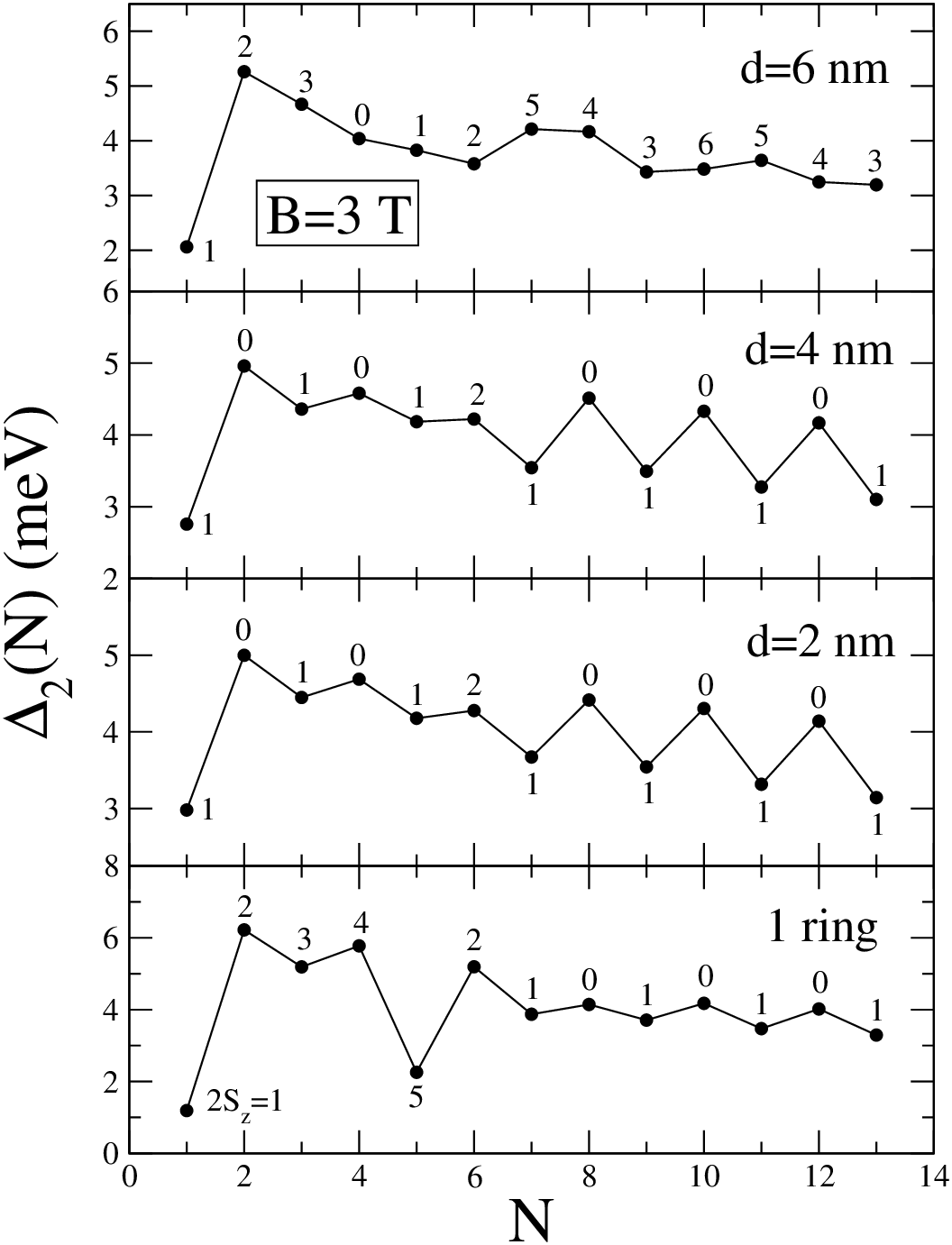} }
\caption{
Same as Fig. \ref{fig9} for $B=3$ T.
}
\label{fig10}
\end{figure}

\begin{figure}[f]
\centerline{\includegraphics[width=10cm,clip]{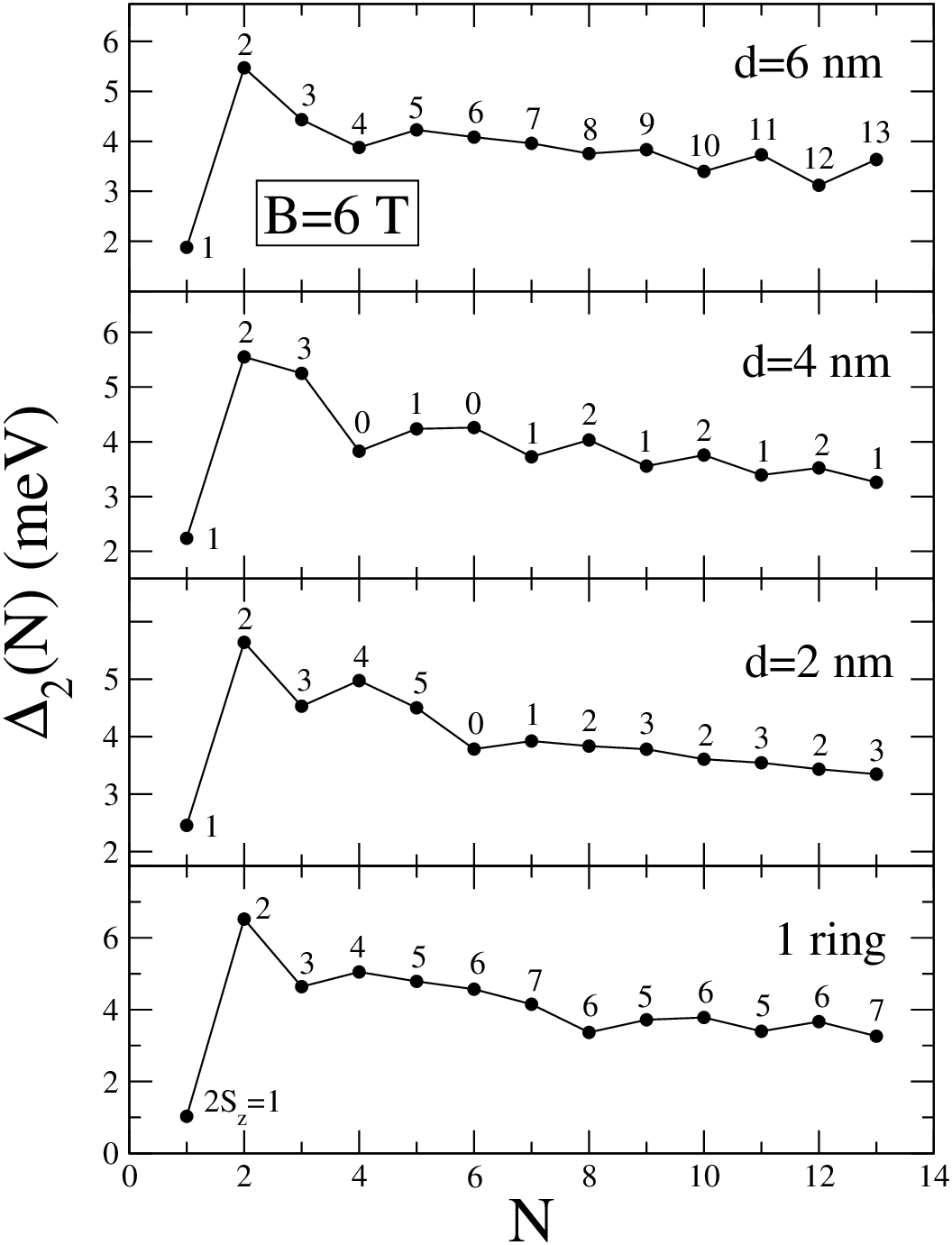} }
\caption{
Same as Fig. \ref{fig9} for $B=6$ T.
}
\label{fig11}
\end{figure}

We have also calculated the addition energies, defined by
\begin{equation}
\Delta_2(N)= E(N+1)- 2 E(N)+ E(N-1)  \;\;\; ,
\end{equation}
$E(N)$ being the total energy of the $N$-electron system, for QRMs made of up to
14 electrons at different inter-ring distances, submitted to several
magnetic fields, as a function of $N$. For the sake of comparison, we have also 
calculated $\Delta_2(N)$ for the corresponding single rings. The results for
$B=0, 3,$ and 6 T are shown in Figs. \ref{fig9}-\ref{fig11}, respectively, in
which the bottom panels correspond to the single ring.

From Fig. \ref{fig9} one can see that at zero magnetic field the single-QR addition
spectrum presents the usual intense peaks at $N=2, 6$ and 10 with zero total
spin, and those at $N=4$ and 8 with $S_z=1$ satisfying Hund's rule.
Similar results are found for the QRMs with $d=2$ and 4 nm, indicating that
such systems behave as a single ring owing to the strong quantum mechanical
coupling corresponding to these inter-ring distances --notice that the spin
values coincide for all the configurations but that with $N=13$. This fact contrasts
with the results found for the vertical ring molecules of Ref. \onlinecite{Mal06},
where at $d=4$ nm the spectrum clearly reflected an intermediate coupling situation
due to the filling of the first antibonding orbitals. As commented before, for the
systems studied in the present paper such states are only occupied for larger
inter-ring separations (or $N$'s of the order of 30).

The spectrum corresponding to $d=6$ nm is shown in the top panel of the same 
figure. One can see that, although some of the marked peaks are preserved, in
particular those at $N=2$ and 8, the ones at $N=4,6$ no longer
exist --notice that for 6 electrons the spectrum presents now a minimum
and also that a new peak is found at $N=5$.
This intricate structure can be understood from the corresponding 
single-particle energy levels. Indeed, it appears that the QRMs with
$N\leq 4$ are made up of only bonding states, the first antibonding state being 
filled when $N=5$.
From $N\geq 7$ on, the QRMs have always occupied both B and AB orbitals but, 
however, the intermediate 6-electron configuration has again only symmetric states.
This alternate behavior evidences that 6 nm is not a separation large enough for 
the QRMs to be in the weak coupling limit, but rather corresponds to an intermediate regime.
Notice also that, from the results of Ref. \onlinecite{Mal06}, in the weak coupling limit 
one would expect to find clearly marked peaks at the same $N$ values as for the
single ring multiplied by two --i.e. at $N=4$, 12 and 20, indicating that the rings
are so apart that behave as
isolated entities. We have checked that for our QRMs to present such spectrum, we
should consider inter-ring distances of about 10 nm.
The different spin values for $d=6$ nm as compared to those in the strong coupling
regime can also be explained from the sp levels. For example, the $2S_z=3$ assignation
of the QRM with $N=5$ is due to the above-mentioned filling of an antibonding
--spin-up with $l=0$-- orbital replacing the spin-down $|l|=1$ state occupied
for $d=2$ and 4 nm.
Analogously, the configuration with $S_z=1$ (instead of $S_z=0$) for $N=10$
can also be explained from the sp levels: in the strong coupling limit,
the QRM is formed by the spin-degenerated sp levels with $l=0,|1|$ and $|2|$, but
this closed-shell configuration is prevented by the filling of the antisymmetric
orbitals at $d=6$ nm. Finally, the reverse situation occurs at $N=8$, where
the closing of the antibonding $l=0$ and $|1|$ shells contrasts with the Hund's-rule
configurations found for the strongly coupled molecules.

Fig. \ref{fig10} shows the addition energies corresponding to the situation in which
a magnetic field of 3 T is applied to the rings. 
Like what is found at $B=0$, the spectrum of the single system and of the molecules with
$d=2$ and 4 nm are rather similar --notice the different energy scales,
the most remarkable difference being the salient minimum that appears for the single
QR at $N=5$.
For the above-mentioned inter-ring distances, peaks with $S_z=0$ are found at 
$N=2,4,8,10$ and 12, as well as a peak at $N=6$ with $S_z=2$, although they
are not as clearly marked as at $B=0$.
It turns out that, even in the presence of a magnetic field, when the
single-particle energy levels no longer display the $\pm l-$degeneration,
the QRMs can adopt configurations that are somehow analogue to these
characteristic of the situation at $B=0$, namely the closed-shell
ones and those fulfilling Hund's rule. Indeed, for e.g. $d=4$ nm and $N=10$,
the ring molecule is made up of the spin-degenerate bonding states with $l=0-4$
(instead of those with $|l|=0-2$ of the $B=0$ case). Similarly, at $N=6$
the occupied orbitals are the spin-up and -down ones with $l=1$ and 2, and the
spin-up ones with $l=0$ and 3 (instead of the spin-degenerate states with
$|l|=0-1$ filled at zero magnetic field).
For larger inter-ring separations, the occupancy of the first antibonding 
orbitals washes out these structures and the addition spectrum becomes
flatter and irregular.
One can notice also the different spin assignations between the single and 
the coupled systems, especially for the lowest-populated configurations. In
particular, the single QRs with $N\leq 5$ turn out to be fully spin-polarized, 
which can be attributed to the combined effect of the magnetic field and a 
relatively strong exchange-correlation interaction characteristic of few-electron 
single quantum rings. The relatively higher spin values at $d=6$ nm for $N\geq7$ 
are due to the filling of the antibonding states.

Finally, the addition energies for $B=6$ T are shown in Fig. \ref{fig11}. It can be
seen that in all cases the only clearly marked peak is the one at $N=2$, with the
rest of the spectra being rather flat, following the trend observed at $B=3$ T.
Nevertheless, some weak peaks are still found and can be interpretated 
as in the previous cases, e.g. the one at $N=8$ for
$d=4$ nm with $2S_z=2$: the system fills the spin-up and -down
states with $l=2-4$ and the spin-up ones with $l=1$ and 5.
One can also notice that the faint peak of the 4-electron configurations 
of both the single ring and the QRM with $d=2$ nm becomes a minimum 
at larger inter-ring distances.
Concerning the spin, the single QRs and the QRMs with $d=2$ and 4 nm turn 
out to be fully polarized for $N\leq 7,5$ and 3, respectively, whereas the 
filling of the antibonding states favors the fully spin-polarization of 
molecules with the largest ring separation for all the considered electron 
numbers.

\section{Summary}

Within the local spin-density functional theory, we have addressed the
ground state of quantum ring molecules containing up to 40 electrons,
with different inter-ring distances, and submitted to perpendicular
magnetic fields. In the strong coupling regime the energy levels and 
the addition energies of the QRMs are similar to those of a single QR,
although some differences are found due to the effect of the magnetic 
field, which has a tendency to wash out the clearly marked peaks 
characteristic of the $B=0$ case as well as to yield flatter addition spectra.
However, even at $B\neq 0$, some peaks are still present and they can be
interpretated as at zero magnetic field.

When the ring separation is increased until the first antibonding orbitals
are occupied, the addition spectra become irregular and the ring molecules
are fully spin-polarized at relatively low magnetic fields.
The filling of such states yields isospin oscillations as a function of $B$, 
increasing in one-unit jumps once the corresponding molecular configurations
reach the maximum spin value.

Despite the lack of experimental results to compare ours with, we believe 
that the ones herewith presented may be helpful in the analysis of future 
experiments on vertically coupled QRs concerning, e.g., the realization of 
single-electron transistor (SET) measurements, where the evolution of the 
chemical potential $\mu(N)$ with the magnetic field can be experimentally
identified as the variation of the position of the current peaks as a function 
of the applied field, showing irregularities arising from phase transitions.

\section*{ACKNOWLEDGMENTS}
This work has been performed under grants FIS2005-01414 from DGI (Spain),
Generalitat Valenciana FPI (MR), and projects  UJI-Bancaixa
P1-1B2006-03 (Spain) and 2005SGR00343 from Generalitat de Catalunya.




\begin{thebibliography}{99}

\bibitem{Jac98} L. Jacak, P. Hawrylak, and A. Wójs, \emph{Quantum Dots} (Springer, Berlin, 1998);
T. Chakraborty, \emph{Quantum Dots}, (Elsevier, Amsterdam, 1999); D. Bimberg, M. Grundmann, N.N.
Ledentsov, \emph{Quantum Dot Heterostructures}, (Wiley, Londres, 2001).

\bibitem{Lip08} E. Lipparini, \emph{Modern Many-Particle Physics} (World Scientific, Singapore, 2008).

\bibitem{Tar96} S. Tarucha, D.G. Austing, T. Honda,  R.J. van
der Hage, and L.P. Kouwenhoven, Phys. Rev. Lett. {\bf 77}, 3613 (1996);

\bibitem{Gar97} J. M. Garc\'{\i}a, G. Medeiros-Ribeiro, K. Schmidt, T. Ngo,
J. L. Feng, A. Lorke, J. Kotthaus,
and P. M. Petroff, Appl. Phys.  Lett. {\bf 71}, 2014 (1997).

\bibitem{Lor00} A. Lorke, R.J. Luyken, A.O. Govorov, J.P. Kotthaus, J.M. Garcia,
and P.M. Petroff, Phys. Rev. Lett. {\bf 84}, 2223 (2000).

\bibitem{Fuh01} A. Fuhrer, S. L\"uscher, T. Ihn, T. Heinzel,
K. Ensslin, W. Wegscheider, and M. Bichler, Nature {\bf 413}, 822
(2001);  T. Ihn, A. Fuhrer, T. Heinzel, K. Ensslin, W.
Wegscheider, and M. Bichler, Physica E {\bf 16}, 83 (2003).

\bibitem{Hel99} R. Held, S. L\"uscher, T. Heinzel, K. Ensslin, and W. Wegscheider,
Appl. Phys. Lett. {\bf 75}, 1134 (1999).

\bibitem{Gon05} Z. Gong, Z.C. Niu, S.S. Huang, Z.D. Fang, B.Q. Sun, and J.B. Xia, Appl.
Phys. Lett {\bf 87}, 093116 (2005).

\bibitem{Kle07} N. A. J. M. Kleemans, I. M. A. Bominaar-Silkens, V. M.
Fomin, V. N. Gladilin, D. Granados, A. G. Taboada, J. M. Garc\'{\i}a,
P. Offermans, U. Zeitler, P. C. M. Christianen, J. C. Maan, J. T.
Devreese, and P. M. Koenraad, Phys. Rev. Lett. {\bf 99}, 146808 (2007).

\bibitem{Ihn05} T. Ihn, A. Fuhrer, K. Ensslin, W. Wegscheider,
and M. Bichler, Physica E {\bf 26}, 225 (2005).


\bibitem{Cha94} T. Chakraborty and P. Pietil\"ainen,
Phys. Rev. B {\bf 50}, 8460 (1994).

\bibitem{Kri94}
I.V. Krive , R.I. Shekhter, S.M. Girvin, and M. Jonson, 
Physica Scripta {\bf T54}, 123 (1994);

\bibitem{Emp99}
A. Emperador, M. Barranco, E. Lipparini, M. Pi, and Ll. Serra,
Phys. Rev. B {\bf 59}, 15301 (1999).

\bibitem{Lin01}
J.C. Lin and G.Y. Guo, Phys. Rev. B {\bf 65}, 035304 (2001).

\bibitem{Aic06}
M. Aichinger, S.A. Chin, E. Krotscheck, and E. R\"as\"anen,
Phys. Rev. B {\bf 73}, 195310 (2006).

\bibitem{Liu08}
Y.M. Liu, G.M. Huang, T.Y. Shi, Phys. Rev. B {\bf 77}, 115311 (2008).

\bibitem{Bli96} R.H. Blick, R.J. Haug, J. Weis, D. Pfannkuche, K.v. Klitzing, and K. Eberl,
Phys. Rev. B {\bf 53}, 7899 (1996).

\bibitem{Sch97} G. Schedelbeck, W. Wegscheider, M. Bichler, and G. Abstreiter,
Science {\bf 278}, 1792 (1997).

\bibitem{Ron01} M. Rontani, F. Troiani, U. Hohenester, and E. Molinari,
Solid State Commun. {\bf 119}, 309 (2001).

\bibitem{Hol02} A.W. Holleitner, R.H. Blick, A.K. H\"{u}ttel, K. Eberl, and J.P. Kotthaus,
Science {\bf 297}, 70 (2002).

\bibitem{Ota05} T. Ota, M. Rontani, S. Tarucha, Y. Nakata, H.Z. Song, T. Miyazawa, T. Usuki,
M. Takatsu, and N. Yokoyama, Phys. Rev. Lett. {\bf 95}, 236801 (2005).

\bibitem{Par00} B. Partoens and F.M. Peeters, Phys. Rev. Lett.
{\bf 84}, 4433 (2000); B. Partoens and F.M. Peeters, Europhys.
Lett. {\bf 56}, 86 (2001).

\bibitem{Pal95} J.J. Palacios and P. Hawrylak, Phys. Rev. B {\bf 51}, 1769  (1995).

\bibitem{Anc03} F. Ancilotto, D.G. Austing, M. Barranco, R. Mayol, K. Muraki, M. Pi,
S. Sasaki, and S. Tarucha, Phys. Rev. B {\bf 67}, 205311 (2003).

\bibitem{Aus04} D.G. Austing, S. Tarucha, H. Tamura, K. Muraki, F.
Ancilotto, M. Barranco, A. Emperador, R. Mayol, and M. Pi,
Phys. Rev. B {\bf 70}, 045324 (2004).

\bibitem{Bel06} D. Bellucci, M. Rontani, G. Goldoni, and E. Molinari,
Phys. Rev. B {\bf 74}, 035331 (2006).

\bibitem{Man05} T. Mano, T. Kuroda, S. Sanguinetti, T. Ochiai, T. Tateno,
J. Kim, T. Noda, M. Kawabe, K. Sakoda, G. Kido and N. Koguchi,
Nanoletters {\bf 5}, 425 (2005).

\bibitem{Kur05} T. Kuroda, T. Mano, T. Ochiai, S. Sanguinetti, K.
Sakoda, G. Kido, and N. Koguchi, Phys. Rev. B {\bf 72}, 205301 (2005).

\bibitem{Sua04} F. Su\'arez, D. Granados, M.L. Dotor, and J.M.
Garc\'{\i}a, Nanotechnology {\bf 15}, S126 (2004).

\bibitem{Gra05} D. Granados, J.M. Garc\'{\i}a, T. Ben, and S.I. Molina,
Appl. Phys. Lett. {\bf 86}, 071918 (2005).

\bibitem{Ahn00} K.-H. Ahn and P. Fulde,
Phys. Rev. B {\bf 62}, R4813  (2000).

\bibitem{Li04} Y. Li and H. M. Lu, Jpn. J. Appl. Phys. {\bf 43}, 2104 (2004).

\bibitem{Dia07} L. G. G. V. Dias da Silva, J. M. Villas-B\^oas, and S. E. Ulloa,
Phys. Rev. B {\bf 76}, 155306 (2007).

\bibitem{Sza07} B. Szafran, S. Bednarek, and M. Dudziak,
Phys. Rev. B {\bf 75}, 235323  (2007).

\bibitem{Sza08} B. Szafran, Phys. Rev. B {\bf 77}, 235314  (2008).

\bibitem{Pia07} G. Piacente and G.-Q. Hai,
J. Appl. Phys. {\bf 101}, 124308 (2007).

\bibitem{Cli05}
J. I. Climente and J. Planelles, Phys. Rev. B {\bf 72}, 155322 (2005).

\bibitem{Cas06} L. K. Castelano, G.-Q Hai, B. Partoens, and F. M.
Peeters, Phys. Rev. B {\bf 74}, 045313  (2006).

\bibitem{Mal06} F. Malet, M. Barranco, E. Lipparini, R. Mayol, M. Pi,
J. I. Climente, and J. Planelles, Phys. Rev. B {\bf 73}, 245324  (2006).

\bibitem{Fer94} M. Ferconi and G. Vignale, Phys. Rev. B {\bf 50}, 14722 (1994).

\bibitem{Ron99} M. Rontani, F. Rossi, F. Manghi, and E. Molinari, Solid
State Commun. {\bf 112}, 151 (1999);
M. Rontani, S. Amaha, K. Muraki, F. Manghi, E. Molinari, S. Tarucha,
and D.G. Austing, Phys. Rev. B. {\bf 69}, 085327 (2004).

\bibitem{Li01} S.S. Li, and J.B. Xia, J. Appl. Phys. {\bf 89}, 3434
(2001); A. Puente, and Ll. Serra, Phys. Rev. B {\bf 63}, 125334 (2001);
J.I. Climente, J. Planelles, and F. Rajadell, J. Phys.: Condens.
Matter {\bf 17}, 1573 (2005).

\bibitem{Pi01} M. Pi, A. Emperador, M. Barranco, and F. Garcias,
Phys. Rev. B {\bf 63}, 115316 (2001).

\bibitem{Lun83} S. Lundqvist,
{\em Theory of the Inhomogenous
Electron Gas}, edited by S. Lundqvist and N. H. March
(Plenum, New York, 1983) p. 149.

\bibitem{Per81} J. P. Perdew and A. Zunger,
Phys. Rev. B {\bf 23}, 5048 (1981).


\newpage

\end{thebibliography}
\end{document}